# Simple model for frequency response of a resonant tunneling diode caused by potential change of quantum well due to electron charge

Masahiro Asada[1)] and Safumi Suzuki

Department of Electrical and Electronic Engineering, Tokyo Institute of Technology,

Meguro-ku, Tokyo, Japan.

1) E-mail: asada@pe.titech.ac.jp

Abstract:

The frequency dependence of negative differential conductance (NDC) is an important property for the resonant-tunneling-diode terahertz source. Among several phenomena determining the frequency dependence, this paper shows that the effect of potential change of the quantum well due to electron charge can be analyzed with a simple and tractable model based on the tunneling admittance and capacitance. The result is identical to that of Feiginov's analysis based on more fundamental equations, showing a one-to-one correspondence between the parameters of the two analyses. Similar to Feiginov's analysis, our analysis also shows that NDC remains finite even at infinitely high frequency. It is shown in our model that this result is attributed to neglecting the tunneling time at the emitter barrier. Comprehensive analysis of the frequency dependence of NDC will be possible by incorporating the tunneling time into the present model.

The terahertz (THz) frequency band (roughly 0.1-10 THz) has a possibility of various applications, such as imaging, spectroscopy, high-capacity communications, radars, and so on [1][2]. A compact THz source is an important component for these applications, and the resonant-tunneling-diode (RTD) oscillator is a candidate for such a THz source [3]-[7].

In the RTD THz oscillators, it is important to investigate the frequency dependence of the negative differential conductance (NDC). Several intrinsic factors are responsible for the frequency dependence of NDC [8]-[18], such as the tunneling and dwell times in the resonant tunneling region, transit time in the collector depletion region, and potential change of the quantum well due to electron charge, in addition to parasitic elements around RTD [3]. Feiginov has reported theoretical analysis of the frequency dependence caused by the dwell time in the quantum well and the potential change of the quantum well due to electron charge [15]. Comparison between the theory and experiment at low frequency has also been reported [15] [19].

In this paper, we analyze the frequency dependence of NDC caused by these phenomena with a simple and tractable model based on the tunneling admittance and capacitance of RTD, instead of Feiginov's analysis based on more fundamental equations of the tunneling and potential. The parameters used in our model are found to have a one-to-one correspondence to those in Feiginov's analysis. NDC is shown to remain finite even at infinitely high frequency, similar to Feiginov's analysis. In our model, this result is attributed to neglecting the tunneling time at the emitter barrier. Comprehensive analysis will be possible by incorporating the tunneling time into the present model.

Figure 1 shows the potential profile of RTD with applied voltage. $J_e$ and $J_c$ in Fig.1 are the electron flow densities (tunneling electron densities per unit time and unit area) from emitter to quantum well and from quantum well to collector, respectively, which have an opposite direction to the current densities. $N_e$ is the carrier density per unit area at the interface between the emitter and the emitter barrier, and $N_{qw}$ is that in the quantum well. Only one resonance level is assumed in the quantum well. The carrier density in the collector region and its back flow to the quantum well are neglected, although generalization including them is possible. $V_e$ is the applied voltages across the emitter barrier together with the emitter accumulation layer and left half of the quantum well, and $V_c$ is that across the collector barrier together with the collector depletion layer and right half of the quantum well. $e$ is the electron charge. The collector spacer layer and the transit time in this layer the collector depletion layer [16] is neglected in this analysis.

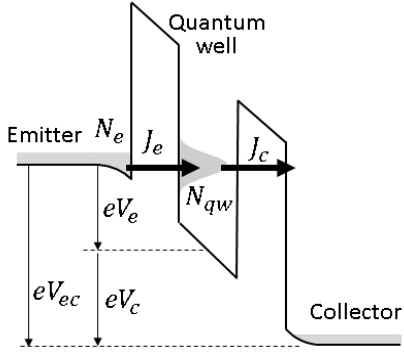

**Fig. 1** Potential profile and electron flows with applied voltage in RTD.

The basic equations are composed of those for the electron flow densities and the potential change. The equations for the electron flow densities are given as follows in the sequential-tunneling model.

$$J_e = T_e(N_e - N_{qw}), \quad (1)$$

$$J_c = T_c N_{qw}, \quad (2)$$

$$\frac{dN_{qw}}{dt} = J_e - J_c, \quad (3)$$

where $T_e$ and $T_c$ are the tunneling transmission rates for $J_e$ and $J_c$, respectively. The potential change is discussed with an equivalent circuit including capacitances instead of the basic equation of the potential, as shown later.

The steady-state ($d/dt = 0$) solutions are obtained from Eqs. (1)–(3) as

$$J_e = J_c = \frac{T_e T_c}{T_e + T_c} N_e, \quad (4)$$

$$N_{qw} = \frac{T_e}{T_e + T_c} N_e. \quad (5)$$

The differential conductance per unit area at the emitter barrier is calculated as

$$G_e = \frac{\partial J_e}{\partial V_e} = \frac{T_c}{T_e + T_c}\left(\frac{T'_e T_c}{T_e + T_c} N_e + T_e N'_e\right), \quad (6)$$

where $T'_e = \partial T_e/\partial V_e$ and $N'_e = \partial N_e/\partial V_e$. The electron charge $e$ is omitted in Eq. (6). In the NDC region of the current-voltage curve of RTD, $G_e < 0$ due to $T'_e < 0$. The total NDC for the applied voltage across RTD is proportional to $G_e$, as shown later. The total NDC is not simply obtained by $G_e$ multiplied by the ratio of the total thickness to the emitter barrier thickness because of the potential change of the quantum well due to electron charge.

By a small AC voltage superimposed on the bias voltage, a small AC term is added to each parameter in Fig. 1, e.g., $V_e$ is expressed as $V_e + \mathrm{Re}[v_e e^{j\omega t}]$, where we use the complex representation with the imaginary unit $j$, Re[ ] denotes real part in the square brackets, $v_e$ and $\omega$ are the amplitude and angular frequency of the AC voltage, respectively, and $V_e$ in this expression is the DC component due to the bias voltage. Hereafter, other parameters in Fig. 1 are expressed in the same manner: DC and AC components in uppercase and lowercase, respectively. We assume the small-signal case (DC components ≫ AC components) in the present analysis.

The basic equations for the AC components of the electron flow densities and electron density in the quantum well, which are denoted with lowercase as mentioned above, are obtained from Eqs. (1)–(3) as

$$j_e = T'_e(N_e - N_{qw})v_e + T_e(n_e - n_{qw}), \quad (7)$$

$$j_c = T'_c N_{qw} v_c + T_c n_{qw} \simeq T_c n_{qw}, \quad (8)$$

$$j\omega n_{qw} = j_e - j_c. \quad (9)$$

In the last part of Eq. (8), the effect of $T'_c = \partial T_c / \partial V_c$ is small and neglected. This is equivalent to neglecting the conductance $G_c = \partial J_c / \partial V_c$ ($>0$) per unit area. Generalization of the analysis including $G_c$ is possible. Assuming that electrons are supplied from the emitter electrode to the emitter accumulation layer much faster than the response of RTD, we may approximate $n_e \simeq N'_e v_e$.

From Eqs. (5)–(9), we obtain

$$j_e = Y_e v_e, \quad (10)$$

$$j_c = \frac{Y_e}{1 + j\omega \tau_c} v_e, \quad (11)$$

where

$$Y_e = \frac{1 + j\omega \tau_c}{1 + j\omega \tau_{dwell}} G_e, \quad (12)$$

$$\tau_c = \frac{1}{T_c}, \quad (13)$$

$$\tau_{dwell} = \frac{1}{T_e + T_c}. \quad (14)$$

Equations (10) and (11) are expressed by an equivalent circuit, as shown in Fig. 2(a). The capacitances $C_e$ and $C_c$ of the emitter and collector barrier regions per unit area are added in Fig. 2(a). The potential change of the quantum well due to electron charge can be considered with these capacitances.

Figure 2(a) is reduced to Fig. 2(b). The admittance $Y_{RTD} + j\omega C_{ec}$ in Fig. 2(b) is given by $(j_e + j_{de})/(v_e + v_c)$ in Fig. 2(a), which is deformed to $(Y_e + j\omega C_e)/(1 + v_c/v_e)$ by dividing the numerator and denominator by $v_e$. The ratio $v_c/v_e$ in this equation is calculated with the relation $j_e + j_{ed} = j_c + j_{cd}$ in Fig. 2(a) as $v_c/v_e = [\tau_c Y_e/(1 + j\omega\tau_c) + C_e]/C_c$. By this process, $Y_{RTD}$ is calculated as

$$Y_{RTD} = G_\infty + \frac{G_0 - G_\infty}{1 + j\omega\tau_{resp}}, \quad (15)$$

where

$$G_0 = G_e \frac{C_c}{\tau_c G_e + C_e + C_c}, \quad (16)$$

$$G_\infty = G_e \left(\frac{C_c}{C_e + C_c}\right)^2 \frac{\tau_c}{\tau_{dwell}}, \quad (17)$$

$$\tau_{resp} = \tau_{dwell} \frac{C_e + C_c}{\tau_c G_e + C_e + C_c}. \quad (18)$$

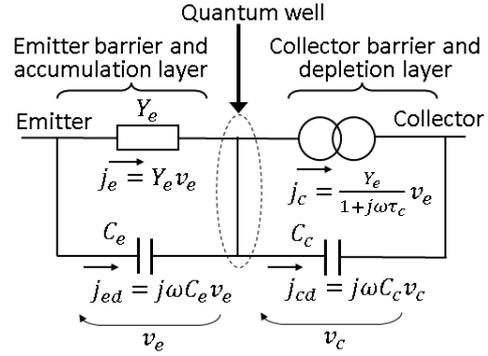

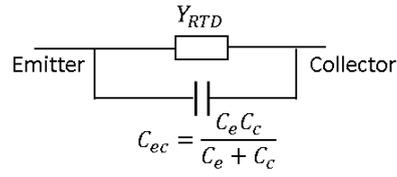

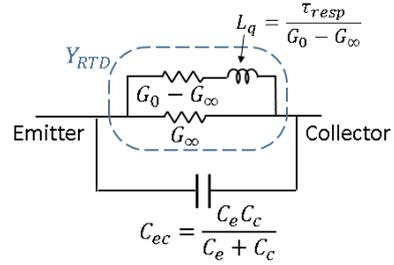

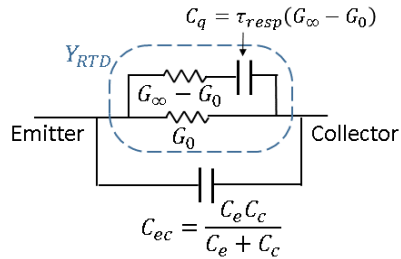

**Fig. 2** Equivalent circuits of RTD. (a) is resulted from Eqs. (10) and (11) with capacitances per unit area, $C_e$ and $C_c$, for potential change due to electron charge. Differential conductance at collector barrier is neglected. (a) is reduced to (b) which is expressed as (c) and (d) with $Y_{RTD}$ in Eq. (15). (c) and (d) are equivalent to each other.

$G_0$ and $G_\infty$ in Eqs. (16) and (17) are the conductances of RTD at $\omega = 0$ and $\infty$, respectively, and $\tau_{resp}$ in Eq. (18) is the response time of $Y_{RTD}$ in Eq. (15). $\tau_{resp}$ is also the response time of $n_{qw}$.

Equations (15)–(18) are just equivalent to the result in Feiginov's analysis [15] for the case neglecting the change in the collector barrier height with applied voltage ($v'_c = 0$ in [15]). Our parameters have a one-to-one correspondence with those in [15] as $T_e \to v_e$, $T_c \to v_c$, $C_e \to \varepsilon/d$, $C_c \to \varepsilon/l = C_{wc}$, $C_e + C_c \to C$, $N_{qw} \to N_{2D}$, and $N'_e \to \rho_{2D}$. (Parameters in [15] indicated on the right sides of the above arrows represent the same physical quantities as those in the present model on the left sides.) The equivalence holds even when the change in the collector barrier height is considered.

From Eq. (18), the response time $\tau_{resp}$ of RTD is longer than the electron dwell time $\tau_{dwell}$ in the quantum well for the NDC region of the current-voltage curve where $G_e < 0$.

Figure 2(b) is expressed as (c) and (d) with Eq. (15), which are equivalent to each other. Fig. 2(c) is the same as that in [15]. The inductance $L_q$ in Fig. 2(c) is negative in the NDC region of the current-voltage curve, while the capacitance $C_q$ in Fig. 2(d) is positive in the NDC region.

Equation (16) implies that the effect of potential change of the quantum well due to electron charge is included in the first term of the denominator, $\tau_c G_e$. If this factor is neglected, the ratio $G_0/G_e$ is equal to the ratio between the total voltage across RTD and the voltage applied to the emitter barrier region. Since $G_e < 0$ in the NDC region, the absolute value of $G_0$ increases due to the effect of electron charge in the quantum well.

As shown in Eq. (15), NDC (the real part of $Y_{RTD}$ in the NDC region) decreases in absolute value with frequency at low frequency ($\omega\tau_{resp} \lesssim 1$), but remains finite at infinitely high frequency. The finite NDC at infinitely high frequency arises from the neglect of the emitter tunneling time, as discussed below.

In the limit of $\omega \to \infty$, $j_c$ and $n_{qw}$ approach zero, while $j_e$ remains finite, as seen from Eqs. (8), (10), and (11). In this situation, the finite tunneling current through the emitter barrier $j_e$ reaches the collector through the collector displacement current $j_{cd}$, in spite that the tunneling current through the collector barrier $j_c$ vanishes. The origin of the finite $j_e$ at $\omega \to \infty$ is the finite $G_e$ included in Eq. (10) through Eq. (12). As a result, NDC is finite at $\omega \to \infty$ because $G_e$ given by Eq. (6) is frequency independent.

Furthermore, the origin of the frequency-independent $G_e$ is $T'_e$ in Eq. (7). In the present analysis, we assumed that $T'_e$ is frequency independent, i.e., the tunneling time at the emitter barrier is negligibly small. NDC may degrade and approach zero with increasing frequency by the finite tunneling time [21] in addition to the phenomena discussed here, although frequency range in which the finite tunneling time is effective may be higher. The temporal behavior of the tunneling current must be analyzed quantum-mechanically to obtain the finite tunneling time.

In our previous paper [16], we phenomenologically introduced tunneling and dwell times for the resonant tunneling to the discussion of the frequency response. By the analytical tunneling time instead of phenomenological one together with the result of the present paper, the frequency response of NDC can be expressed more accurately.

A large-signal analysis as an extension of Feiginov's model has been reported [20]. The present analysis can also be extended to the large-signal analysis by incorporating the nonlinear voltage dependence of the tunneling rate.

In summary, the frequency response of RTD taking into account the electron dwell time in the quantum well and the potential change of the quantum well due to electron charge was analyzed with simple and tractable model based on tunneling admittance at the emitter barrier and capacitance. The result obtained in this analysis was identical to that of Feiginov's analysis [15], and there is a one-to-one correspondence between the parameters of the two analyses. NDC remaining finite even at infinitely high frequency was attributed to neglecting the tunneling time at the emitter barrier. The analysis in this paper can be a first step to construct a comprehensive analysis including the dwell time in the quantum well, the tunneling time, and the potential change due to the electron charge in the quantum well.

Acknowledgement

The authors thank Honorary Prof. Y. Suematsu, Emeritus Profs. K. Furuya and S. Arai, Profs. Y. Miyamoto and N. Nishiyama, and Assoc. Prof. M. Watanabe of the Tokyo Institute of Technology for their continuous encouragement. This study was supported by a scientific grant-in-aid (21H04552) from JSPS, CREST (JPMJCR21C4) from JST, X-NICS (JPJ011438) from MEXT, SCOPE (JP215003005) from MIC, commissioned research from NICT (No. 03001), and the Canon Foundation.

References

[1] M. Tonouchi, Nature Photonics **1**, 97 (2007).
[2] T. Nagatsuma, IEICE Electron. Express **8**, 1127 (2011)..
[3] E. R. Brown, J. R. Sonderstrom, C. D. Parker, L. J. Mahoney, K. M. Molvar, and T. C. McGill, Appl. Phys. Lett. **58**, 2291 (1991).
[4] M. Reddy, S. C. Martin, A. C. Molnar, R. E. Muller, R. P. Smith, P. H. Siegel, M. J. Mondry, M. J. W. Rodwell, H. Kroemer, S. J. Allen, IEEE Electron Device Lett. **18**, 218 (1997).
[5] M. Feiginov, Int. J. Infrared and Millimeter Waves, **40**, 365 (2019).
[6] M. Asada and S. Suzuki, Sensors, **21**, 1384 (2021).
[7] D. Cimbri, J. Wang, A. Al-Khalidi, and E. Wasige, IEEE Trans. THz Sci. Technol. **12**, 226 (2022).
[8] W. Frensley, Phys. Rev. B, **36**, 1570 (1987).
[9] R. K. Mains and G. I. Haddad, J. Appl. Phys. **64**, 5041 (1988):
[10] N. C. Kluksdahl, M. A. Kriman, D. K. Ferry, Phys. Rev. B, **39**, 7720 (1989).
[11] H. C. Liu, Phys. Rev. B, **43**, 12538 (1991); Erratum, **48**, 4877(1993).
[12] A. Sugimura, Semicond. Sci. Technol. **9**, 512 (1994).
[13] W.-R. Liou and P. Robin, IEEE Electron Dev. **41**, 1098 (1994).
[14] M. Asada, Jpn. J. Appl. Phys. **40**, 5251 (2001).
[15] M. Feiginov, Appl. Phys. Lett. **78**, 3301(2001).
[16] M. Asada, S. Suzuki, and N. Kishimoto, Jpn. J. Appl. Phys. **47**, 4375 (2008).
[17] J.-F. Mennemann, A. Jüngel, and H. Kosina, J. Computational Phys. **239**, 187 (2013).
[18] K. S. Grishakov, V. F. Elesin, M. M. Masov, and K. P. Katin, Adv. Mat. Sci. Eng. vol.2017, 2031631 (2017).
[19] M. Feiginov and D. R. Chowdhury, Appl. Phys. Lett. **91**, 203501 (2007).
[20] P. Ourednik, G. Picco, D. T. Nguyen, C. Spudat, and M. Feiginov, J. Appl. Phys. **133**, 014501 (2023).
[21] E. H. Hauge and J. A. Støvneng, Rev. Mod. Phys., **61**, 917 (1989).